\documentclass[twocolumn,letterpaper]{IEEEAerospaceCLS}  % only supports two-column, letterpaper format

\usepackage[]{graphicx}    % We use this package in this document
\newcommand{\ignore}[1]{}  % {} empty inside = %% comment

\usepackage{amsmath,amssymb,amsfonts}
\usepackage[dvipsnames]{xcolor}
\def\BibTeX{{\rm B\kern-.05em{\sc i\kern-.025em b}\kern-.08em
    T\kern-.1667em\lower.7ex\hbox{E}\kern-.125emX}}

\usepackage{mleftright}

\usepackage{import}
\usepackage{algorithm}
\usepackage{algpseudocode}

\usepackage{tikz,pgfplots,tikz-3dplot}
\usepackage{pgfplots}
\usepackage{textcomp}
\usetikzlibrary{positioning, plotmarks, arrows, calc, shadings, patterns,
	decorations.pathreplacing, calc,shapes,external}

\pgfplotsset{compat=newest}
\usetikzlibrary{plotmarks}
\usetikzlibrary{arrows.meta}
\usepgfplotslibrary{fillbetween}
\usepgfplotslibrary{patchplots}
\usepgfplotslibrary{groupplots}
\usepackage{grffile}
\usepackage{multirow}
\usepackage{array}

\usepackage{url}

\tikzstyle{blockdiag}	= [node distance=5mm, >=stealth', semithick]
\tikzstyle{block}			= [draw, rectangle, minimum width=1cm, minimum 
height=.8cm]
\tikzstyle{sum} = [draw,circle,inner sep=0pt, minimum size=6pt]
\tikzstyle{connector} = [draw,circle,inner sep=0pt, minimum size=2pt, 
fill=black]
\tikzstyle{branch} = [circle,inner sep=0pt,minimum size=1mm,fill=black,draw=black]
\tikzstyle{gain} = [draw,regular polygon, regular polygon 	sides=3,thick,minimum height=3em,minimum width=4em, rotate=30]
\tikzstyle{bguide} = [rectangle,minimum height=3em,minimum	width=4em]
\tikzstyle{line} = [thick]
\tikzstyle{guide} = [anchor=center]
\tikzstyle{input} = [coordinate]
\tikzstyle{output} = [coordinate, node distance=1cm]

\pgfdeclarelayer{background}
\pgfdeclarelayer{foreground}
\pgfsetlayers{background,main,foreground}
\definecolor{mycolor1}{rgb}{0.00000,0.44700,0.74100}%
\definecolor{mycolor2}{rgb}{0.85000,0.32500,0.09800}%
\definecolor{mycolor3}{rgb}{0.92900,0.69400,0.12500}%
\definecolor{mycolor4}{rgb}{0.49400,0.18400,0.55600}%
\definecolor{mycolor5}{rgb}{0.46600,0.67400,0.18800}%

\definecolor{mycolor6}{rgb}{0.00000,0.05263,0.97368}%
\definecolor{mycolor7}{rgb}{0.00000,0.10526,0.94737}%
\definecolor{mycolor8}{rgb}{0.00000,0.15789,0.92105}%
\definecolor{mycolor9}{rgb}{0.00000,0.21053,0.89474}%
\definecolor{mycolor10}{rgb}{0.00000,0.26316,0.86842}%
\definecolor{mycolor11}{rgb}{0.00000,0.31579,0.84211}%
\definecolor{mycolor12}{rgb}{0.00000,0.36842,0.81579}%
\definecolor{mycolor13}{rgb}{0.00000,0.42105,0.78947}%
\definecolor{mycolor14}{rgb}{0.00000,0.47368,0.76316}%
\definecolor{mycolor15}{rgb}{0.00000,0.52632,0.73684}%
\definecolor{mycolor16}{rgb}{0.00000,0.57895,0.71053}%
\definecolor{mycolor17}{rgb}{0.00000,0.63158,0.68421}%
\definecolor{mycolor18}{rgb}{0.00000,0.68421,0.65789}%
\definecolor{mycolor19}{rgb}{0.00000,0.73684,0.63158}%
\definecolor{mycolor20}{rgb}{0.00000,0.78947,0.60526}%
\definecolor{mycolor21}{rgb}{0.00000,0.84211,0.57895}%
\definecolor{mycolor22}{rgb}{0.00000,0.89474,0.55263}%
\definecolor{mycolor23}{rgb}{0.00000,0.94737,0.52632}%
\definecolor{mycolor24}{rgb}{0.00000,1.00000,0.50000}%

\usepackage{array}% in the preamble
\newcommand{\norm}[1]{\left\|#1\right\|}
\newcommand{\abs}[1]{\left|#1\right|}
\newcommand{\field}[1]{\mathbb{#1}}
\newcommand{\R}{{\field{R}}}

\newcommand{\bmtx}{\begin{bmatrix}}
\newcommand{\emtx}{\end{bmatrix}}
\newcommand{\bsmtx}{\left[ \begin{smallmatrix}} 
\newcommand{\esmtx}{\end{smallmatrix} \right]} 
\newcommand{\bmatarray}[1]{\left[\begin{array}{#1}}
\newcommand{\ematarray}{\end{array}\right]} 

\definecolor{blue1}{RGB}{222,235,247}
\definecolor{blue2}{RGB}{158,202,225}
\definecolor{blue3}{RGB}{49,130,189}

\definecolor{mycolor1}{rgb}{0.00000,0.44700,0.74100}%
\definecolor{mycolor2}{rgb}{0.85000,0.32500,0.09800}%
\definecolor{mycolor3}{rgb}{0.92900,0.69400,0.12500}%
\definecolor{mycolor4}{rgb}{0.49400,0.18400,0.55600}%
\definecolor{mycolor5}{rgb}{0.46600,0.67400,0.18800}%

\newtheorem{mytheo}{Theorem}

\newcommand{\edtn}[1]{\null} %{ {\color{red} #1} } {\null}

\graphicspath{{figures/}}

\begin{document}
\title{Linear Parameter Varying Attitude Control For CubeSats Using Electrospray Thrusters}

\author{%
Felix Biertümpfel \\
Technische Universität Dresden\\
Helmholtzstr. 10 \\
01069 Dresden, Germany\\
felix.biertuempfel@tu-dresden.de
\and 
Emily Burgin\\ 
Technische Universität Dresden\\
Helmholtzstr. 10 \\
01069 Dresden, Germany\\
emily.burgin@tu-dresden.de
\and 
Harald Pfifer\\ 
Technische Universität Dresden\\
Helmholtzstr. 10 \\
01069 Dresden, Germany\\
harald.pfifer@tu-dresden.de
\and
Hanna-Lee Harjono\\
Massachusetts Institute of Technology\\
77 Massachusetts Avenue \\
Cambridge, MA 02139\\
hharjono@alum.mit.edu\\
\and
Paulo Lozano\\ 
Massachusetts Institute of Technology\\
77 Massachusetts Avenue \\
Cambridge, MA 02139\\
plozano@mit.edu
%%%% IMPORTANT: Use the correct copyright information--IEEE, Crown, or U.S. government. %%%%%
\thanks{\footnotesize 979-8-3503-5597-0/25/$\$31.00$ \copyright2025 IEEE}              % This creates the copyright info that is the correct 2025 data.
%\thanks{{U.S. Government work not protected by U.S. copyright}}         % Use this copyright notice only if you are employed by the U.S. Government.
%\thanks{{979-8-3503-5597-0/25/$\$31.00$ \copyright2025 Crown}}          % Use this copyright notice only if you are employed by a crown government (e.g., Canada, UK, Australia).
%\thanks{{979-8-3503-5597-0/25/$\$31.00$ \copyright2025 European Union}}    % Use this copyright notice is you are employed by the European Union.
}

\maketitle

\thispagestyle{plain}
\pagestyle{plain}

\maketitle

\thispagestyle{plain}
\pagestyle{plain}

\begin{abstract}

This paper proposes the design of a single linear parameter-varying (LPV) controller for the attitude control of CubeSats using electro spray thrusters. CubeSat attitude control based on electro spray thrusters faces two main challenges. Firstly, the thruster can only generate a small control torque leading to easily saturating the actuation system. Secondly, CubeSats need to operate multiple different maneuvers from large to small slews to pointing tasks. LPV control is ideally suitable to address these challenges. The proposed design follows a mixed-sensitivity control scheme.  The parameter-varying weights depend on the attitude error and are derived from the performance and robustness requirements of individual typical CubeSat maneuvers. The controller is synthesized by minimizing the induced $L_2$-norm of the closed-loop interconnections between the controller and weighted plant. The performance and robustness of the controller is demonstrated on a simulation of the MIT Space Propulsion Lab's Magnetic Levitation CubeSat Testbed.
%This paper proposes the design of a single linear parameter-varying (LPV) controller for the attitude control of CubeSats using electro spray thrusters. CubeSat attitude control based on electro spray thrusters faces two main challenges. Firstly, the thruster can only generate a small control torque leading to easily saturating the actuation system. Secondly, CubeSats need to operate multiple different maneuvers from large to small slews to pointing tasks. LPV control is ideally suitable to address these challenges. The proposed design follows a mixed-sensitivity control scheme with LPV weights that are derived from the performance and robustness requirements of individual typical CubeSat maneuvers. The controller is synthesized by minimizing the induced $L_2$-norm of the closed-loop interconnections between the controller and weighted plant. The performance and robustness of the controller is demonstrated on a simulation of the MIT Space Propulsion Lab's Magnetic Levitation CubeSat Testbed.
\end{abstract}

\tableofcontents

%%%%%%%%%%%%%%%%%%%%%%%%%%%%%%%%%%%%%%
\section{Introduction}
%%%%%%%%%%%%%%%%%%%%%%%%%%%%%%%%%%%%%%

CubeSats provide a cost-effective way for scientific investigations and in-orbit technology demonstrations. However, current CubeSat propulsion and attitude control systems limit their mission spectrum due to energy requirements as well as limited performance, lifetime, and reliability \cite{Quinsac2020,Li2013}.

Novel ionic electrospray engines (iESE) for CubeSats can overcome these technical limitations \cite{Lozano2015}. Electrospray engines are a type of electric spacecraft propulsion that work by emitting positively or negatively charged
particles from an electrically conductive liquid using a strong electric field. They are passively fed, compact, fuel-efficient, and can be used for main propulsion or combined attitude and position control~\cite{MierHicks2017,GIANNIPECORA2023}. 
Due to their small form factor, multiple stages of iESE can be installed on a CubeSat and then be sequentially used~\cite{JiaRichards2020}. Staging achieves longer missions, higher reliability, and thrust than standard CubeSat propulsion. Moreover, iESEs are completely throttleable and jitter-free. Hence, using electrospray thrusters for attitude and position control facilitates new levels of pointing performance for CubeSats which cannot be achieved by classical reaction wheels \cite{Addari2017,Douglas2021}. CubeSats equipped with staged ionic electrospray engines can thus pave the way to more complex and ambitious missions such as space telescopes and interferometers composed of one or multiple CubeSats, free-flying coronagraphs, or CubeSat swarms for space debris mitigation. For example, space telescopes built by an iESE CubeSat constellation can be re-orientated or even moved to different orbits, and reach high pointing accuracies~\cite{GomezJenkins2018}. A high-performing and robust CubeSat control system is a key component for all these missions. The CubeSats require highly precise attitude and position control to follow various guidance profiles. At the same time, the control system must robustly handle a multitude of missions and unplanned abrupt maneuvers for, e.g., collision avoidance to fulfill the stringent requirements for space operations despite the low torque available from iESE ~\cite{NietoPeroy2019} . 

This paper proposes a robust linear parameter varying (LPV) attitude control design relying exclusively on iESEs as actuators for complex CubeSat operations. 
The control design is rooted in the induced $L_2$-framework following a mixed sensitivity approach~\cite{wu1996induced}. LPV control was successfully applied for attitude control for a large spacecraft, in order to account for varying system dynamics over the orbit and mission~\cite{Burgin2023}. The designed controller provides inherent performance and robustness guarantees mandatory for complex space missions. Recently LPV control was proposed to handle different control requirements for different modes in spacecraft operation by using parameter-varying weights, see~\cite{Burgin2025}. Similarly, an ad-hoc scheduled approach without the guaranteed performance was proposed in~\cite{Lurie2000} for single axis attitude control. The present paper builds upon these ideas by designing the CubeSat controller with weights depending  on the difference between the current and final target attitude.
The resulting controller is then scheduled with this attitude difference. Large attitude changes can be controlled slower so as not to saturate the iESE, whereas small changes and disturbance rejection during pointing operation can be faster. This allows, for example, the implementation of a single controller covering both large attitude slew maneuvers and fine pointing. Thus, dedicated slewing and pointing controllers and required switching between them can be avoided~\cite{Geshnizjani2013Thesis,Rufus2002,Biannic2010}. The result is a smoother transient and simplified implementation. Moreover, using LPV control provides guaranteed robustness and performance across the parameter domain. Thus, stability is guaranteed as the spacecraft transitions from maneuvering to pointing

The proposed LPV controller is designed for the Magnetic Levitation CubeSat Testbed (MagLev) - a magnetically levitated testbed floating a 1U mockup CubeSat using magnetic fields inside a vacuum chamber. MagLev is designed by MIT's Space Propulsion Laboratory (SPL)  specifically for attitude control experiments of iESE driven CubeSat control systems~\cite{MierHicks2017diss}. The focus is primarily on different size attitude maneuvers, also referred to as slew maneuvers, followed by a substantial time at a pointing task where the satellite has to maintain a given absolute attitude in space~\cite{MierHicks2017}. The developed controller should avoid thrust saturation as much as possible during nominal maneuvers. 
Given the limited resources of CubeSat operations, the controller does not necessarily follow smooth precalculated trajectories that are typically used for large satellites or compex missions, see, e.g.,~\cite{Mclnnes1998,Longuski2013,Marshall2023}. The CubeSat may also needs to follow step or ramp-like changes in attitude. Thus, the controller must provide necessary disturbance rejection capabilities even during sudden large commanded attitude changes.

The paper is structured as follows. After an introduction, Section~\ref{sec:LPVctrl} presents a background on linear parameter varying control and mixed sensitivity design. Section~\ref{sec:CtrlProb} provides a description of the MagLev testbed including a model of its dynamics and the setup for the control problem. In Section~\ref{sec:SlewCtrl}, the LPV control design is described in detail. Section~\ref{sec:Eval} concludes the paper with an extensive controller evaluation in a nonlinear simulation environment.

%
%\
%
%Alternative Chapter Structure following the launcher paper
%\begin{enumerate}
%	\item Background on Robust Control, if we do LPV then it is necessary
%	\item Electrospray Thruster CubeSat Control Design Problem
%	\begin{enumerate}
%		\item CubeSat Model (rigid, flexible, thrusters, disturbance)
%		\item Mission Problem (Example missions, Control Objectives)
%	\end{enumerate}
%	\item Eletctrospray Thruster Control Design Problem
%\end{enumerate}

\section{Background}\label{sec:LPVctrl}

\subsection{Linear Parameter-Varying Systems}

LPV systems are a class of systems whose state-space matrices depend continuously on a time-varying parameter vector $\rho: \R^+ \rightarrow \mathcal{P}$. The compact subset $\mathcal{P} \in $$\R$$^{n_\rho}$ is selected based on physical considerations. In addition, the parameter variation rates $\dot{\rho}$ are confined to lie in a hyper-rectangle $\dot{\mathcal{P}}$ defined by $\dot{\mathcal{P}}= \{\dot{\rho}(t) \in \,\R^{n_\rho} | \abs{\dot{\rho}_i(t)} \leq \nu_i, i=1,\hdots,n_\rho   \}$. Hence, the set of all admissible trajectories is $\mathcal{T} = \{ \rho: \R \rightarrow {\mathcal{P}}\, | \, \rho \in $${\mathcal{C}}$$^1, \rho(t) \in \mathcal{P}$ and $\dot{\rho}(t) \in \dot{\mathcal{P}} \, \forall t \geq 0\}$. 
A general LPV system $P_\rho$ is given by:
\begin{equation}\label{eqn:lpvsystem}
\bmtx \dot{x}(t) \\ y(t) \emtx \!\!=\!\! \bmtx  A(\rho(t)) & B(\rho(t))  \\ C(\rho(t)) & D(\rho(t))\emtx \bmtx x(t) \\ u(t) \emtx\!,
\end{equation}
where $x(t) \in \R^{n_x}$ is the state vector, $u(t) \in \R^{n_u}$ the input vector, and $y(t) \in \R^{n_y}$ the output vector.  The state space matrices are continuous functions of the parameter vector with appropriate dimensions, e.g., $A: \mathcal{P} \rightarrow $${\R}$$^{n_x \times n_x}$. In the remainder of the paper the explicit time-dependence is mostly omitted when clear from context.

The performance of an LPV system can be specified in terms of its induced $L_2$-norm from input $u$ to output $y$: 
\begin{equation}\label{eqn:inducedL2norm}
	\norm{P_\rho} = \sup_{u \in {L}_2 \setminus \{0\}, \rho \in {\mathcal{T}}, x(0)=0} \frac{\norm{y}_2}{\norm{u}_2}.
\end{equation}
A generalization of the Bounded Real Lemma in \cite{wu1996induced} states a sufficient condition to upper bound $\norm{P_\rho}$, which is given in Theorem~\ref{theo:genBRL}.
\begin{mytheo}\label{theo:genBRL}
	\cite{wu1996induced}:	 $G_\rho$ is exponentially stable and $\norm{G_\rho} \leq \gamma$ if there exists a continuously differentiable symmetric matrix function $X: \mathcal{P} \rightarrow \R$$^{n_x \times n_x}$ such that $X(p) \geq 0$ and
	\begin{equation} \label{eq:brl}
		\bmtx XA+A^TX + \partial X & XB \\ B^TX & -I \emtx + \frac{1}{\gamma^2} \bmtx C^T \\ D^T \emtx \bmtx C & D \emtx \leq 0
	\end{equation}
	hold for all $p \in \mathcal{P}$ and $q \in \dot{\mathcal{P}}$, where $\partial X$ is defined as $\partial X(p,q) = \sum_{i=1}^{n_\rho} \frac{\partial X}{\partial \rho_i}(p) q_i$. In (\ref{eq:brl}), the dependence of the matrices on $p$ and $q$ has been omitted to shorten the notation.
\end{mytheo}
%\begin{figure}[h!bt] 
%\centering
%\input{figures/obs_control_design_MixedSensitivity_in}
%\caption{Four-block mixed sensitivity problem.}
%\label{fig:obs_control_design_MixedSensitivity}
%\end{figure}

This theorem extends to the induced $L_2$-norm controller synthesis in Wu, et al., \cite{wu1996induced}. Consider an open-loop LPV system $G_\rho$ with the state-space formulation as in (\ref{eqn:lpvsystem}) with inputs denoted $[w^T, u^T]^T$ and outputs $[z^T, y^T]^T$, where $w$ and $z$ are measures of performance. The objective is to synthesize a controller $K_\rho$,
\begin{equation}\label{eqn:K}
	K_\rho:	\bmtx \dot{x}_K \\ u \emtx = \bmtx A_K(\rho) & B_K(\rho) \\ C_K(\rho) & D_K(\rho) \emtx \bmtx x_K \\ y \emtx,
\end{equation}
such that the induced $L_2$-gain of the closed-loop interconnection of $G_\rho$ and $K_\rho$, denoted by the lower fractional transformation $F_l(G_\rho,K_\rho)$, is minimized.
\begin{equation}\label{eqn:minimisation}
	\min_{K_\rho} \norm{F_l(G_\rho,K_\rho)}.
\end{equation}
Thus, the optimization of the performance of the closed-loop system can be solved via parametrized LMI conditions; see \cite{wu1996induced} for details. This synthesis problem involves an infinite collection of LMI constraints parametrized by $(p,q) \in \mathcal{P} \times \dot{\mathcal{P}}$. A remedy to this infinite dimensionality is to approximate the constraints with finite-dimensional LMIs evaluated on a gridded domain of $p$ and $q$. Tools to solve the synthesis problem are readily available; \texttt{LPVTools} \cite{hjartarson2015lpvtools} is used in this paper.

\subsection{Linear Parameter-Varying Mixed-Sensitivity Design}\label{sec:LPVmixedsens}

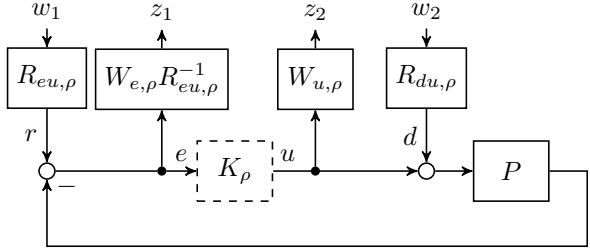
\begin{figure}[h!] 
	\centering
	\begin{tikzpicture}[blockdiag, auto]
	
	% Blocks
	\node[block] (Plant) {$P$};
	\node[sum, left=of Plant] (SumP) {};
	\node[branch, left=of SumP,xshift=-8mm] (Branch1) {};
	\node[block, above=of Branch1,yshift=0.25cm] (Wu) {$W_{u,\rho}$};
	
	\node[block, dashed,left=of Branch1] (Controller) {$K_\rho$};	
	\node[branch, left=of Controller, xshift=0.1cm] (BranchE) {};
	\node[sum, left=of BranchE, xshift=-0.85cm] (SumE) {};
	\node[block, above=of BranchE,yshift=0.25cm] (We) {$W_{e,\rho}R_{eu,\rho}^{-1}$};
	\node[block, above=of SumE,yshift=0.21cm] (W1) {$R_{eu,\rho}$};
	
	\node[block, above=of SumP,yshift=0.21cm] (Wd) {$R_{du,\rho}$};
	% 
	% 
	% \node[block, right=of Wd, xshift = 1.4cm] (Weta) {$W_z$};
	% \node[block, right=of Weta] (Wy) {$W_y$};
	% \node[block, above=of Sum2] (Wn) {$W_2$};
	
	% % Conncections
	%\draw[->] (Pd) -| (SumP);
	\draw[<-] (Plant) -- (SumP);
	\draw[<-] (SumE) -- (W1)node[pos = 0.5, left]{$r$};
	\draw[<-] (W1.north) -- +(-0,0.25cm) node[above]{$w_1$};
	\draw[<-] (Wd.north) -- +(0,0.25cm) node[above]{$w_2$};
	\draw[->] (We.north) -- +(0, +.25cm) node[above]{$z_1$};
	\draw[->] (Wu.north) -- +(0, +.25cm) node[above]{$z_2$};
	
	\draw[-] (SumE) -- (BranchE);
	\draw[->] (BranchE) -- (Controller) node[pos=0.5] {$e$};
	\draw[->] (BranchE) -- (We);
	\draw[->] (Controller) -- (SumP) node[pos=0.1] {$u$};
	\draw[->] (Branch1) -- (Wu);
	\draw[->] (Wd) -- (SumP)node[pos = 0.5, left]{$d$};

	\draw[->] ($(Plant.east)$) -| +(+.5cm,-1.0cm) -| (SumE.south) node[pos=0.95,swap] {$-$};
	% \draw[->] ($(Plant.south east)!.3!(Plant.north east)$) +(+.8cm,-0.0cm) node[branch] {} -| (Wy);
	% \draw[<-] (Wn.north) -- +(0, +.5cm)node[left, name=d2]{$w_2$};
	% \draw[<-] (Wd.north) -- +(0, +.5cm)node[left, name=d1]{$w_1$};
	% 
	% \draw[->] (Wn.south) -- (Sum2.north);
	% \draw[->] (Wd.south) -- (Sum1.north);
	% \draw[<-] (Wu.south) -- (Con1.north);
	% \draw[->] (Sum2.east) node[below right]{} -- (Controller.west);
	% \draw[->] (Controller.east) -- (Sum1.west) node[pos=0.2]{$u$};
	% \draw[->] (Sum1.east) -- (Plant.west);
	
	\end{tikzpicture} 
	\caption{LPV weighted four-block mixed-sensitivity problem.}
	\label{fig:mixedSens}
\end{figure}

It is common practice to design induced $L_2$-norm optimal controllers by mixed-sensitivity loopshaping; see, e.g., \cite{Skogestad2005}, as recently demonstrated in spacecraft control \cite{Burgin2023}. Consider the closed-loop feedback system between a plant $P$ and controller $K_\rho$ (\ref{eqn:K}). Desired closed-loop behavior can be enforced by minimizing the induced $L_2$-norm of the interconnection between the controller and a weighted, generalized plant $G_\rho$ constructed from $P$ and some weights. The weights are responsible for defining the additional performance in/outputs $w$ and $z$. The proposed weighting scheme applied in this paper uses a minimal number of physically interpretable LPV weights that are derived from the robustness and performance requirements of the closed-loop with respect to the scheduling parameter $\rho$. The scheme is shown in Fig.~\ref{fig:mixedSens}. Note that the subscript $\rho$ indicates a system's dependence on the scheduling parameter $\rho$. Throughout this paper, the plant $P$ is assumed to be linear time-invariant (LTI) as only the performance weights change during typical CubeSat maneuvers. 

Defining the output sensitivity function $S_\rho = (I+P K_\rho)^{-1}$, the generalized closed-loop $F_l(G_\rho,K_\rho)$ of the weighted mixed-sensitivity problem is then
\begin{equation} \label{eqn:mixedSens}
	\bmtx z_1 \\ z_2 \emtx \!\! =\!\! \bsmtx W_{e,\rho} R_{eu,\rho}^{-1} & 0  \\ 0 & W_{u,\rho} \esmtx \!\! \bsmtx S_\rho & -S_\rho P \\ K_\rho S_\rho & -K_\rho S_\rho P \esmtx \!\!\bsmtx R_{eu,\rho} & 0 \\ 0 & R_{du,\rho} \esmtx\!\! \bmtx w_1 \\ w_2 \emtx
\end{equation}
where $W_{e,\rho}$ and $W_{u,\rho}$ denote dynamic parameter-varying weights and $R_{eu,\rho}$, and $R_{du,\rho}$ parameter-dependent scaling factors. The central block is referred to as the \textit{four-block problem} and defines four unique closed-loop mappings that are shaped by the design weights. These four blocks fully describe the performance and robustness of the controlled system. A high magnitude in $W_{e,\rho}$ reduces $S_\rho$ leading to better tracking and disturbance rejection capabilities. A high magnitude in $W_{u,\rho}$ reduces the control effort $K_\rho S_\rho$. Hence, $W_{u,\rho}$ can enforce controller roll-off at high frequencies. The scaling factors are used as the main tuning knobs and are mutually dependent. The  scaling $R_{eu}$ tunes the desired relationship between command signal $u$ and error $e$. A good initial value is the ratio of allowable tracking error to maximum actuator command, implying that the synthesized controller will command its maximum capacity when the tracking error is about to be violated. Similarly $R_{du}$ defines the relationship between expected disturbance $d$ and actuator response $u$. As a result, a third relationship must be considered $R_{de} = R_{du}R_{eu}^{-1}$ which tunes the error $e$ relative to the expected disturbance or in other words, the disturbance rejection performance.

\section{CubeSat Control Design Problem}\label{sec:CtrlProb}
The controller is designed for the Magnetic Levitation CubeSat Testbed (MagLev) device of MIT's Space Propulsion Laboratory (SPL) \cite{MierHicks2017}.
SPL’s MagLev enables direct angular attitude control experiments to
be performed. Testing attitude control capabilities of satellite systems is usually
conducted on air bearings \cite{Schwartz2003}. However, attitude control capability of electrospray
thrusters cannot be evaluated in this manner since they can only operate in vacuum.
Thus, MagLev remedies this since it is based in a vacuum chamber and can achieve
no-friction rotation without air.

\subsection{Magnetic Levitation CubeSat Testbed}
\begin{figure}[h!]		
\centering\includegraphics[width=1.0\columnwidth]{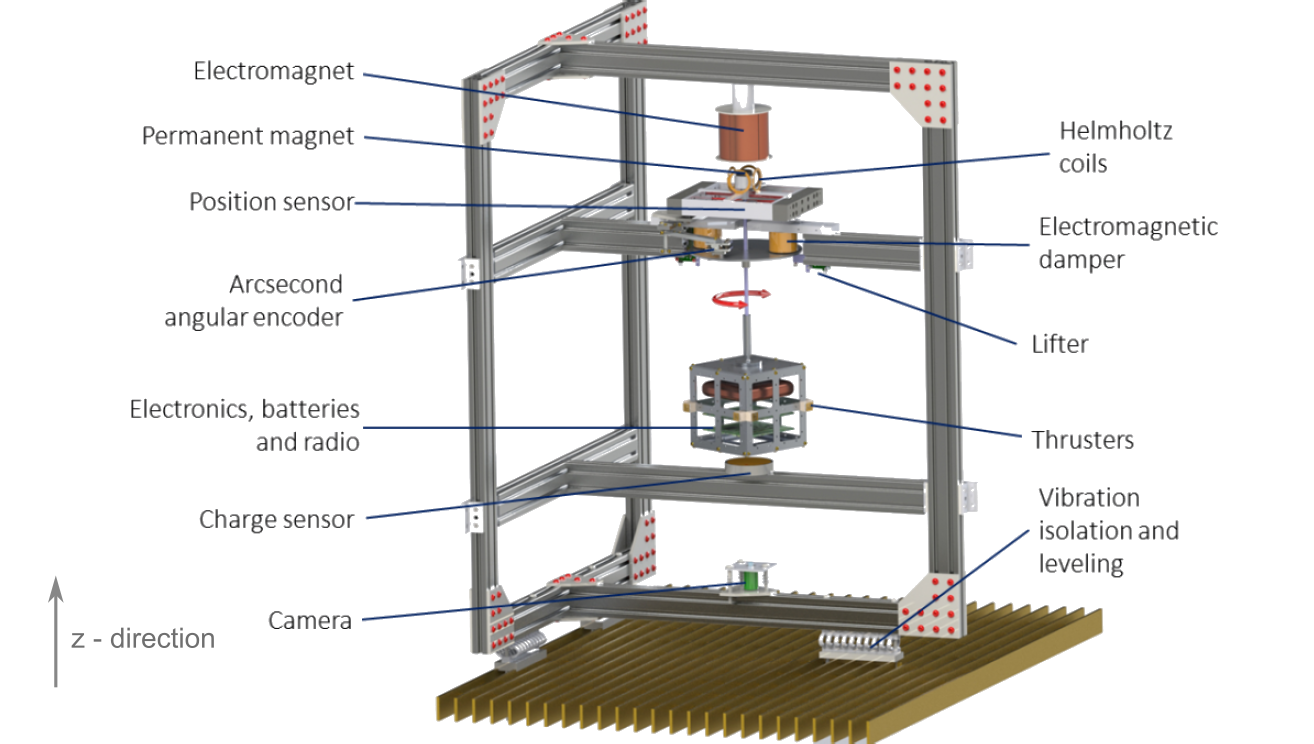}
%\vspace{-5.5pt}
\caption{Diagram of MagLev with main components labeled}
\label{fig:MagDia}
\end{figure}

\begin{figure}[h!]		
\centering\includegraphics[height=0.33\textwidth]{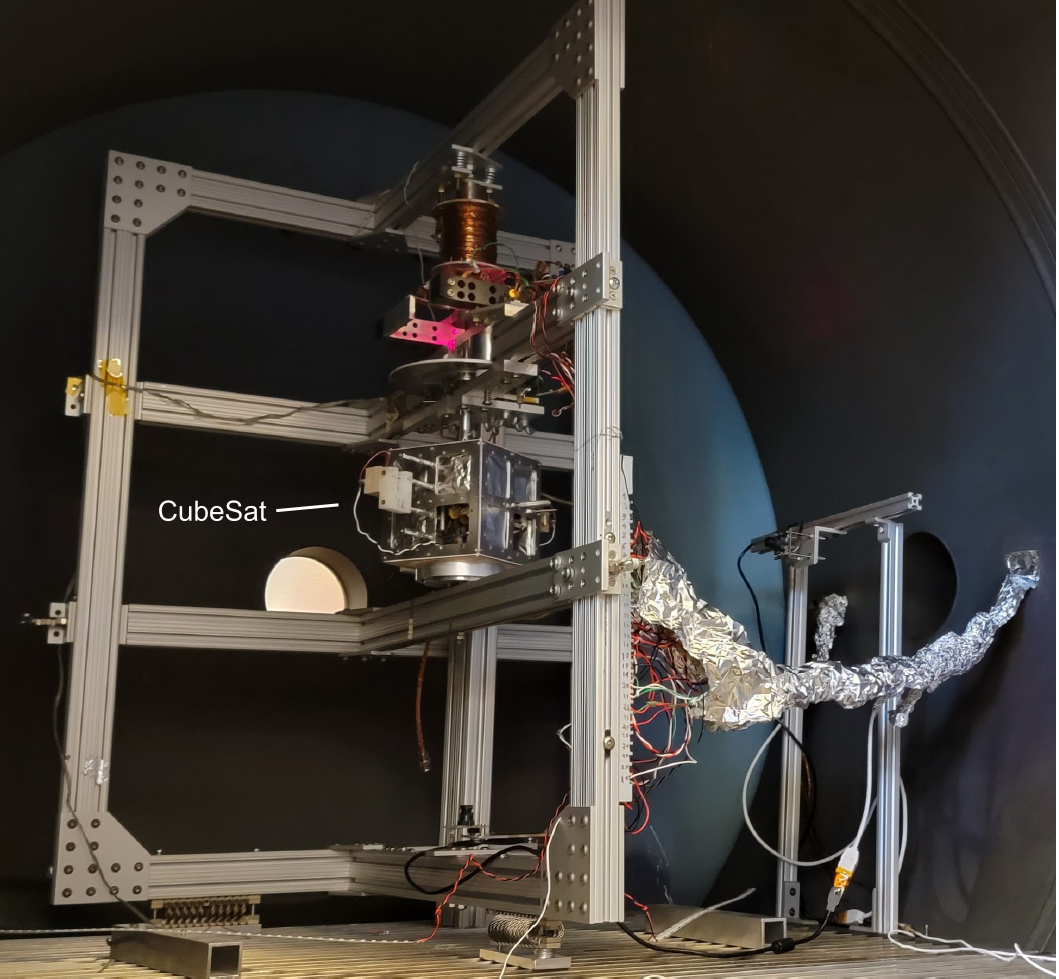}
%\vspace{-5.5pt}
\caption{MagLev device inside the SPL AstroVac vacuum chamber}
\label{fig:MagFoto}
\end{figure}

\begin{figure}[h!]		
\centering\includegraphics[width=0.6\columnwidth]{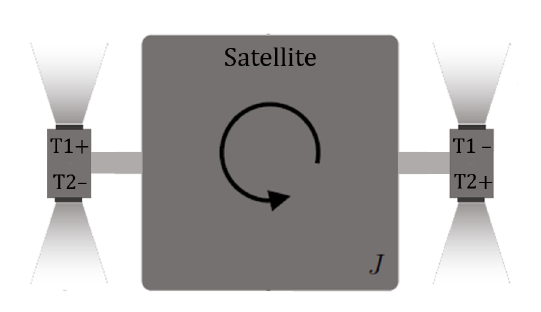}
%\vspace{-5.5pt}
\caption{Top-down view in axis of rotation demonstrating MagLev model satellite
setup with two pairs of thrusters mounted}
\label{fig:Thruster}
\end{figure}

MagLev interfaces with a 1U model CubeSat via magnetic levitation,
enabling $360$ degrees of rotational, 1 degree-of-freedom (1-DOF), zero-friction rotation
about one axis ($z$-axis). Reflecting in-space operation, the levitation mechanically and
electrically isolates the satellite, which is equipped with batteries, a custom PPU, a
radio, and a microcontroller. Angular position is measured with no-contact sensors
- in the current MagLev iteration, with a camera tracking fiducial markers printed
on the bottom of the satellite. In this way, direct thrust measurements can be derived,
along with angular control performance about one axis of rotation. Fig. \ref{fig:MagDia}
shows a diagram of MagLev with some of its critical components highlighted, and
Fig. \ref{fig:MagFoto} shows a picture of MagLev.
Electrospray thrusters are mounted on the MagLev satellite as shown in Fig. \ref{fig:Thruster}. Thrusters are mounted with their thrust vectors orthogonal to the
axis of rotation, on arms extending from the center of opposite satellite faces. They
are mounted and operated in pairs in order to mitigate spacecraft charging effects.
Thus, one thruster pair would apply thrust in a positive rotational direction, and the
other would apply thrust in the negative rotational direction. If only one direction
of rotation is desired, only one thruster pair needs to be mounted, but two pairs are
necessary for full directional control about the free-rotation axis. A thruster pair can
provide $30\,\mu$N thrust, which for the designed lever arm $r$ of $0.21\,$m results
in a maximum torque of $\tau_\text{max}=6.3\,\mu$Nm. The CubeSat's mass moment of 
inertia is $J = 0.006\,\text{kgm}^2$
The CubeSat's rigid body 1-DOF rotational motion about its $z$-axis is given by:
\begin{equation}\label{eq:EoM}
J\ddot{\theta} = \tau +\tau_\text{d}
\end{equation}
where $\ddot{\theta}$ denotes the rotational acceleration, $\tau$ the applied thrust, and $\tau_\text{d}$ an external disturbance torque.

Note that small perturbing torques inherent to the magnetic levitation scheme affect the MagLev
satellite. These torques cause the levitated structure to experience undesired,
long-period rotational oscillations about the 1-DOF axis. The primary contribution
was determined to be caused by magnetic dipole interaction between the levitation
electromagnet and the permanent magnet of the levitated structure. This is undesirable
since the presence of disturbance torques does not allow free-rotation about the
1-DOF axis. To cancel out the torque-producing horizontal magnetic field, Helmholtz
coils were incorporated into the setup. Positioned orthogonally, these Helmholtz coils
enable generation of a magnetic field in any horizontal direction. This Helmholtz
magnetic field can then be used to directly cancel out the field of one of the magnetic
dipoles, effectively preventing dipole interactions and the resulting perturbing torque.
Furthermore, it can also be used to induce angle-dependent torques on the levitating
structure. In this way, user-defined disturbance torques can be applied to the satellite model,
and an attitude controller’s response and robustness to disturbance torques can be
appraised. The paper only considers a nominal disturbance torque $\tau_\text{d}$ built from a constant disturbance torque $\tau_0 =2.5\,\mu$Nm and a $\theta$-dependent torque $\tau_\theta  = 1.6\sin{\theta}\,\mu$Nm, i.e., 
\begin{equation}
\tau_\text{d} = 2.5\,\mu\text{Nm} + 1.6\sin{\theta}\,\mu\text{Nm}.
\end{equation}

\section{Parameter-Varying Control for CubeSat Attitude Maneuvers }\label{sec:SlewCtrl}
%\subsection{Controller Set Up}

%%% the whole slewing problem should be clear from the intro or the sections before
%%
%%Recall that CubeSats, in general, suffer from limited control authority. Electrospray thrusters amplify this problem, due to their thrust in the micro-Newton range..
%%Control saturation is a prevalent problem for CubeSat attitude control. During slewing, the problem is usually addressed by integrating anti-windup schemes into the
%%attitude control system see, e.g., \cite{Ofodile2021,AlHemeary2019} or using a form of "bang-bang" control. see, e.g., \cite{Li2016}.
%%However, operating the system at or close to its saturation limit should be avoided as the control system must also be able to counteract external disturbances. Moreover, these
%%controllers may not provide the necessary (fine) pointing performance required for, e.g., observation tasks.
%%Thus, a separate controller is then used for subsequent pointing task. % should be clear from the intro

The paper proposes an LPV controller synthesis to the CubeSat slewing and pointing problem, i.e., commanded large and potentially abrupt attitude changes and keeping
the commanded attitude. The main idea is to provide a single controller which can handle, both,  the slews of different sizes and precise pointing tasks.
The controller calculates a torque command $\tau_\text{cmd}$ such that the satellites attitude $\theta$ follows a pre-defined guidance reference signal $\theta_\text{ref}$ during the slew.
For large satellites, the reference signal $\theta_\text{ref}$ is typically calculated offline in a way to minimize the required time for the attitude change considering the limitations of the attitude control systems, see, e.g. \cite{Longuski2013}. However, for CubeSat operations in dynamic environments providing such optimal and smooth reference signals is not always possible. 
The slewing and pointing operation pose highly different control objectives, requirements and challenges. 
The control system must autonomously account for these difference. 
It is straightforward to synthesize individual mixed sensitivity controllers for each phase using the weighting scheme introduced in Section \ref{sec:LPVctrl} (Fig. \ref{fig:mixedSens}).
Thus, the weights used for the pointing (subscript $\text{p}$) and acquisition phase (subscript $\text{a}$) can be used to design an LPV controller
which uses these weights at the respective end points of its domain. 

\subsection{Scheduling Parameter}
\begin{figure}
		
			\centering
				\begin{tikzpicture}
	
\begin{axis}[%
	width=1.5in,
	height=1.1in,
	at={(0in,1.2in)},
	scale only axis,
	xmin=0,
	xmax=1000,
	ymin=1,
	ymax=5.5,
	ylabel style={font=\color{white!15!black}},
	ylabel={Magnitude},
	axis background/.style={fill=white},
	axis x line*=bottom,
	axis y line*=left,
	xlabel = {Time},
	title ={\textcolor{white}{|}},
	legend style = {draw = none},
	ymajorticks = false,
	yticklabel = \empty,
	xtick = {112.8},
	xticklabel = {$T$},
	]

\addplot [color=black, line width = 1pt]
table[row sep=crcr]{%
	5.5	4.71238898038473\\
	67.2	4.70820436648035\\
	84.4	4.69565052476742\\
	101.6	4.67472745524594\\
	118.8	4.64543515791581\\
	136	4.60777363277703\\
	153.2	4.56174287982958\\
	170.4	4.50734289907359\\
	187.6	4.44457369050906\\
	204.8	4.37343525413576\\
	222	4.29392758995391\\
	239.2	4.20605069796352\\
	256.4	4.10980457816447\\
	273.6	4.00518923055677\\
	290.8	3.8922046551404\\
	308	3.7708508519155\\
	325.2	3.64112782088193\\
	342.5	3.50224497598356\\
	427.7	2.80527981937485\\
	444.9	2.67626557619485\\
	462.1	2.55562056082351\\
	479.3	2.44334477326083\\
	496.5	2.3394382135067\\
	513.7	2.24390088156122\\
	530.9	2.15673277742428\\
	548.1	2.077933901096\\
	565.3	2.00750425257638\\
	582.5	1.9454438318653\\
	599.7	1.89175263896288\\
	616.9	1.84643067386901\\
	634.1	1.8094779365839\\
	651.3	1.78089442710723\\
	668.5	1.76068014543932\\
	685.7	1.74883509157996\\
	703	1.74532925199435\\
	1005.6	1.74532925199435\\
};\label{pl:thref}

%\addlegendentry{$|\theta_\text{ref}|$}

\addplot [color=Apricot, line width = 1pt]
table[row sep=crcr]{%
	5.5	5.05486677646161\\
	65.7	5.0506828980898\\
	81.4	5.03813126297428\\
	97.1	5.01721187111502\\
	112.8	4.98792472251205\\
	128.5	4.95026981716546\\
	144.2	4.90424715507515\\
	159.9	4.84985673624112\\
	175.6	4.78709856066337\\
	191.3	4.715972628342\\
	207	4.63647893927691\\
	222.7	4.5486174934681\\
	238.4	4.45238829091568\\
	254.1	4.34779133161942\\
	269.8	4.23482661557955\\
	285.5	4.11349414279596\\
	301.2	3.98379391326876\\
	316.9	3.84572592699783\\
	332.6	3.69929018398307\\
	351.7	3.51184423171242\\
	426.2	2.78031338963456\\
	441.9	2.63779858757675\\
	457.6	2.50365154226267\\
	473.3	2.3778722536922\\
	489	2.26046072186546\\
	504.7	2.15141694678243\\
	520.4	2.05074092844302\\
	536.1	1.95843266684733\\
	551.8	1.87449216199536\\
	567.5	1.79891941388712\\
	583.2	1.7317144225226\\
	598.9	1.67287718790169\\
	614.6	1.6224077100245\\
	630.3	1.58030598889093\\
	646	1.54657202450119\\
	661.7	1.52120581685506\\
	677.4	1.50420736595265\\
	693.1	1.49557667179397\\
	712.1	1.49439510239324\\
%	1005.6	1.49439510239324\\
};\label{pl:th}

%\addlegendentry{$|\theta|$}

\draw [color=Apricot, line width = 1pt] plot [smooth, tension=1] coordinates { (712.1, 1.4944) (1000,1.74533)};

\draw [color=black, <->, line width = 1pt] plot [] coordinates { (112.8,	4.98792472251205) (112.8,1.74532925199435)};
\node [circle,draw = none] at (250, 3) {$\rho(T)$};
\addplot [color=mycolor1, dashed, line width = 1pt] plot [] coordinates { (0,1.74532925199435) (1000,1.74532925199435)};\label{pl:thfinal}
%\addplot[-, red!60, solid, line width = 1.5]coordinates{(0,5.7)(103,5.7)};\label{pl:ControlObj}
		
%\addlegendentry{$|\theta_\text{ref}(t\rightarrow\infty)|$}

%\addlegendentry{$|\theta_\text{ref}(t \righterror \infty)|$}

\end{axis}	
	
\end{tikzpicture}
		\caption{Definition of the scheduling parameter $\rho$: Reference signal $|\theta_\text{ref}|$ (\ref{pl:thref}), current attitude $\theta$ (\ref{pl:th}), final attitude $\theta_\text{ref}(t\rightarrow \infty)$ (\ref{pl:thfinal})}
		\label{fig:rhoDefine}
		\vspace{-10pt}
\end{figure}
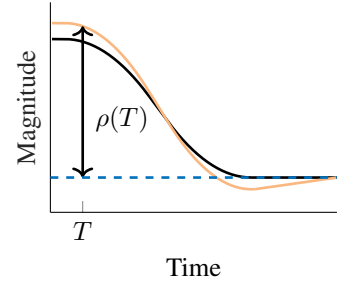

Assume the slewing reference profile ($\theta_\text{ref}$) is provided for a given time-frame leading the CubeSat from its initial attitude $\theta_0$ to the final attitude $\theta_\text{final} = \theta_\text{ref}(t\rightarrow \infty)$. 
Fig.~\ref{fig:rhoDefine} shows that at the beginning of the maneuver the difference between the current attitude $\theta$ and the final attitude $\theta_\text{final}$ is maximal. 
This difference $\Delta \theta =| \theta - \theta_\text{final}|$ will decrease while the CubeSat's attitude converges to its final value by tracking the reference signal.
Once, the CubeSat reaches its final attitude $\Delta \theta$ will remain close to zero. Hence, the controller parameter domain can be defined using $\Delta \theta$ leading to the 
scheduling parameter $\rho = |\theta - \theta_\text{final}|$. This scheduling parameter quantitatively describes the transition from the initial attitude to the final attitude where the pointing task commences. The parameter $\rho$ is confined to lie in the set $[\rho_\text{p}, \rho_\text{a}]$. 
For scheduling values $\rho$ outside of the parameter range,  the controller shall behave as an LTI controller. This is mandatory for pointing as the scheduling parameter refers to the quasi-steady state of the controller. Thus, LTI behavior for $\rho < \rho_\text{p}$ provides the best pointing performance. 
Thus, the parameter bounds $\rho_\text{p}$ and  $\rho_\text{a}$ must be chosen within reasonable proximity to the final and initial attitude error, respectively.
Note that the scheduling parameter only equals the actual tracking error $e_\theta = \theta -\theta_\text{ref}$ for $\theta_\text{ref}(t\rightarrow \infty)$.

\subsection{Pointing Phase Design}

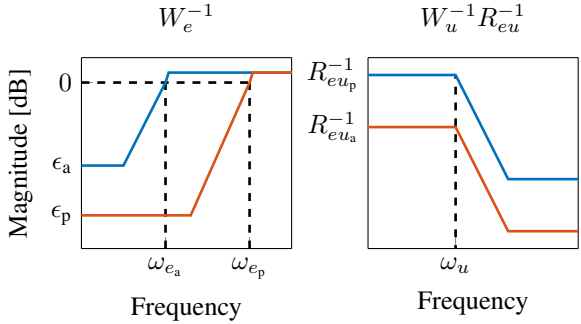
\begin{figure}[t]
		\centering
	\begin{tikzpicture}

\begin{axis}[%
	width=1.1in,
	height=1in,
	at={(0in,0in)},
	scale only axis,
	separate axis lines,
	%every outer x axis line/.append style={white!40!black},
	%every x tick label/.append style={font=\color{white!40!black}},
	%every x tick/.append style={white!40!black},
%	xmode=log,
	xmin=0,
	xmax=2.5,
	xtick = {0,1,2,3},
	xticklabels = {{},{$\omega_{e_\text{a}}$},{$\omega_{e_\text{p}}$},{}}, % normalising
	%every outer y axis line/.append style={white!40!black},
	%every y tick label/.append style={font=\color{white!40!black}},
	%every y tick/.append style={white!40!black},
	ymin=-100,
	ymax=15,
	ytick = {-80, -50, 0},
	yticklabels ={$\epsilon_\text{p}$,$\epsilon_\text{a}$,$0$},
	ytick pos=left,
	xtick pos = bottom,
	ylabel={Magnitude [dB]},
	xlabel={Frequency},
	axis background/.style={fill=white},
	title={$W_e^{-1}$},
	]
	
	\addplot [color=black, dashed, forget plot, line width = 1pt]
	table[row sep=crcr]{%
		2 -100\\
		2  0 \\
	};
	
	\addplot [color=black, dashed, forget plot, line width = 1pt]
	table[row sep=crcr]{%
		1 -100\\
		1 0 \\
	};

		\addplot [color=black, dashed, forget plot, line width = 1pt]
	table[row sep=crcr]{%
		0 0\\
		2 0 \\
	};

\addplot [color=mycolor1, forget plot, line width = 1pt]
table[row sep=crcr]{%
 0  -50\\
0.5 -50 \\
1  0 \\
1.0375 6 \\
3 6 \\
};
\label{line:egacqWeight},

\addplot [color=mycolor2, forget plot, line width = 1pt]
table[row sep=crcr]{%
	0  -80\\
	1.30 -80 \\
	2  0 \\
	2.0375 6 \\
	3 6 \\
};
\label{line:egptWeight},
	
\end{axis}

\begin{axis}[%
	width=1.1in,
	height=1in,
	at={(1.5in,0in)},
	scale only axis,
	separate axis lines,
	%every outer x axis line/.append style={white!40!black},
	%every x tick label/.append style={font=\color{white!40!black}},
	%every x tick/.append style={white!40!black},
	xmin=0.5,
	xmax=3.5,
	%xtick = {0.001,1,100},
	%xticklabels = {$10^0$,$10^1$,$10^2$}, % normalising
	%every outer y axis line/.append style={white!40!black},
	%every y tick label/.append style={font=\color{white!40!black}},
	%every y tick/.append style={white!40!black},
	xtick = {1.75},
	xticklabels={$\omega_{u}$},
	ymin=-70,
	ymax= 40,
	ytick = {0,30},
	yticklabels ={$R_{eu_\text{a}}^{-1}$,$R_{eu_\text{p}}^{-1}$},
	ytick pos=left,
	xtick pos = bottom,
	xlabel={Frequency},
	axis background/.style={fill=white},
	title={$W_u^{-1}R_{eu}^{-1}$},
	legend style={legend cell align=left, align=left, draw=white!15!black}
	]
	
	\addplot [color=black, dashed, forget plot, line width = 1pt]
	table[row sep=crcr]{%
		1.75 -100\\
		1.75  30 \\
	};

		\addplot [color=mycolor1, forget plot, line width = 1pt]
	table[row sep=crcr]{%
		0 30\\
		1.75  30 \\
		2.5 -30 \\
		3.5 -30 \\
	};
	
	\addplot [color=mycolor2, forget plot, line width = 1pt]
	table[row sep=crcr]{%
		0 0 \\
		1.75 0 \\
		2.5 -60 \\
		3.5 -60 \\
	};

\end{axis}
\end{tikzpicture}
	\caption{Frequency domain weights at $\rho_{p}$ (\ref{line:egptWeight}) and $\rho_{a}$ (\ref{line:egacqWeight})}
	\label{fig:LPVWeights}
\vspace{10pt}
\end{figure}

First, the pointing phase of the maneuver is considered. Here, the CubeSat shall maintain a specified orientation in space, with a steady-state error smaller than $\epsilon_\text{p} = 1\cdot 10^{-3}$, i.e., $0.1\%$ error. Accordingly, the shape of $W_e$ must push the sensitivity function $S$ under $\epsilon_\text{p}$ at low frequencies.
$W_e$ is chosen with integral behavior up to $\omega_{e_\text{p}}=0.05\,$ rad/s and a constant value of $0.5$ beyond. Thus, sensitivity
is reduced up to the closed loop bandwidth $\omega_{e_\text{p}}$ and limited to a factor $2\, (6\,\text{dB})$ beyond to limit peak sensitivity and achieve good robustness.
The closed loop bandwidth is chosen low to avoid control saturation.
Fig.~\ref{fig:LPVWeights} depicts the quantitative shape of $W_e$'s frequency response.
Note that the CubeSat has already (double)-integral behavior
 and the steady state error will in theory always eventually vanish. 
The pointing performance is also heavily influenced by the the disturbance rejection $PS$. %where the tuning parameter
%$\epsilon_\text{p} R_{du}R_{eu}^{-1} =\epsilon_\text{p} R_{de}$ refers to the steady state error as result of a disturbance input. 
The selected closed loop bandwidth  $\omega_{e_\text{p}}$ also influences the disturbance rejection capabilities and should be, in general, chosen higher than the encountered low frequency disturbances. For the considered MagLev device these are not critical as in steady state only a constant disturbance occur.
The weight $W_u$ shapes the control sensitivity $KS$ and disturbance control sensitivity $-KSP$, i.e., the actuator response following a reference or disturbance signal, respectively.
$W_u$ is chosen with unit gain up to a given roll-off frequency $\omega_{u_\text{p}} = 10\,$ rad/s and differentiating behavior afterwards. Fig.~\ref{fig:LPVWeights} shows the qualitative shape of $W_u$. Thus controller roll-off for frequencies above $\omega_{u_\text{p}}$ is enforced, which corresponds to the maximum available controller bandwidth. Note that iESP thrusters are extremely fast throtteable and the roll-off is mainly enforced to minimize high frequency firing.
After choosing the dynamic performance weights based on principle system and performance requirements, the ratios $R$ are used to fine tune the pointing controller.
Recall that CubeSats suffer from limited control authority. Electrospray thrusters amplify this problem, due to their thrust in the $\mu$N range. Hence, control saturation is a prevalent problem in the control design. 
$R_{eu}$ directly relates to the control effort and thus the controller gain was chosen to avoid saturation of the electrospray thrusters. The specific value is $R_{eu_\text{p}} = 1/0.75\tau_\text{max}\,$deg/$\mu$Nm, i.e., for a pointing error of $1\,$deg the controller is at $75\%$ of its saturation limit. In a similar fashion, the weight $R_{du,p}$ is chosen such that the thrusters will not saturate under perturbations during operation. The value is $R_{du,p} = 0.05$ which implies that the controller can handle disturbances up to $5\%$ of
the thruster capabilities. The ratio $R_{ed} = R_{eu}R_{du}^{-1}$ is a direct consequence of the other two ratios and refers to disturbance rejection capabilities.
%xxxFB: explaining sentence

%(... non-factor, although would be cool to test).
%Increasing the closed loop bandwidth $\omega_{e,p}$ maximizes the tracking and disturbance rejection capabilities. However, a hard-limitation is the peak sensitivity $\abs{S_\text{max}}$ and sufficient separation between $\omega_{e_\text{p}}$ and $\omega_u$ to satisfy phase margin (PM) requirements. 
\subsection{Slewing Phase Design}
Next, the control requirements for the slewing phase are considered, where the controller tracks a given guidance profile.
 %This can yield less suitable ...
The same tuning knobs are available for the slew controller as for the pointing controller. However, there are three main differences in the slewing design.
First, a larger steady state error $\epsilon_\text{a}$ is allowed, as the CubeSat does not need to tightly track the reference trajectory. Fig.~\ref{fig:LPVWeights} qualitatively displays the reduced tracking. Hence, the $\epsilon_\text{a}$ is larger than $\epsilon_\text{p}$. Specifically, here $\epsilon_\text{p} = 0.1$, which relates to a steady state error of $10\%$.
Second, there needs to be a cautious approach to the tracking bandwidth during this highly dynamic maneuver.
%Recall that CubeSat suffer from limited control authority. Electrospray thrusters amplify this problem, due to their thrust in the micro-Newton range. Hence, control saturation is a prevalent problem in the control design. 
Thus, a lower closed loop bandwidth for the slew controller must be chosen, i.e., $\omega_{e_\text{p}} <\omega_{e_\text{p}}$. For the present example, $\omega_{e_\text{p}} = 0.025\,$rad/s, which is a factor two slower than the pointing closed loop bandwidth. 
The third difference is closely related. Control authority should be limited and smaller than for the pointing phase. This means a smaller value for $R_{eu_\text{a}}$ should be chosen for slewing than for pointing ($R_{eu_\text{p}}$). Another reasoning here is that, the pointing controller will always operate close to the reference signal yielding smaller errors and
necessary compensation. Thus, more control authority can be allowed as the risk of saturations is slightly lower. The selected value for $R_{eu_\text{a}}$ is $1/0.075\tau_\text{max}\,$deg/$\mu$Nm, which is a factor ten higher than for the pointing controller. Thus, the control authority is ten times lower than during pointing to avoid saturation
even for large slew maneuvers.  Increased separation to saturation is also desired to account for time-varying disturbances during the slew. 
The MagLev's characteristic $\theta$ dependent magnetic disturbance torques results in a varying torque disturbance $\tau_\text{d}$ during slew.
Given that the iESP CubeSat uses the same means of control for slewing and pointing and also encounters similar disturbances, $R_{du_\text{a}}$ should be similar to the value of the pointing controller. The final tuning actually yielded $R_{du_\text{a}} = R_{du_\text{p}}$  If more thrusters are used for slewing than for pointing, $R_{du_\text{a}}$ can be increased.

\subsection{Controller Synthesis}
The definition of the the two design points facilitates to pose an LPV induced $L_2$ synthesis problem. During the maneuver, the CubeSat dynamics do not change. Hence, the plant $G_\rho$ is actually linear time-invariant. A chosen function shape, for example, linear, quadratic, or hyperbolic, interpolates the weights $R$ and $W$ over the domain for
specified values of $\rho$. This yields a grid-based LPV controller synthesis with parameter-varying weights. The parameter-varying generalized plant $G_\rho$ takes the form in (\ref{eqn:lpvsystem}). Solving the optimization problem (\ref{eqn:minimisation}) yields the LPV controller of the form (\ref{eqn:K}).
For the present design, a hyperbolic function was chosen so that the rate of change in the controller begins low while the spacecraft is slewing, and then increases to the halfway mark and
decreases again when the CubeSat gets closer to pointing, i.e., when the satellite attitude approaches its final value.
This particular domain shape provided the best results compared to the, also investigated, linear and quadratic domain shapes.
Equation (\ref{eqn:quadDomain}) below provides the definition of $R_{eu}$ as an example.
\begin{subequations}\label{eqn:quadDomain}
	\begin{align}
		R_{eu}(\rho) = & R_{eu_\text{p}} + \alpha_{eu}(1 + \tanh(\rho -\beta_{eu})) \\
		\alpha_{eu} = & \frac{R_{eu_\text{a}}-R_{eu_\text{p}}}{2} \\
		\beta_{eu} = &  \frac{\rho_\text{a}-\rho_\text{p}}{2}+\rho_\text{p}
	\end{align}
\end{subequations}
This formulation guarantees that the end points of the function correspond to the pointing and slewing design points.
The parameter domain is confined to $\rho_\text{p} = 0.01\,$ deg and $\rho_\text{a} = 180\,$ deg. This selection covers the
complete envisioned angular range of slewing maneuvers, while it guarantees that the controller only converges when accurate pointing is reached.
For synthesis, the LPV mixed-sensitivity weights were interpolated across the domain on a grid of $20$ points. 
 \texttt{LPVTools} \cite{hjartarson2015lpvtools} were used to synthesize the parameter-varying controller. The trajectory of $\rho$ in the parameter dependent storage functions was chosen as $p_0 + p_1\rho^2 + p_2\rho^4$, where $p$ are decision variables in the optimization. The rate-bounds of $\dot{\rho}$ were chosen based on ramp shaped slewing signal used for the controller evaluation corresponding to the rates $[-0.1\, 0.1]$ deg/s.
\begin{figure}
	\centering
	\begin{tikzpicture}
\definecolor{blue1}{RGB}{222,235,247}
\definecolor{blue2}{RGB}{158,202,225}
\definecolor{blue3}{RGB}{49,130,189}
%\begin{semilogxaxis}[ width = 0.95\columnwidth, height = 0.6\columnwidth, % Grösse
%  	grid=major, 
%   grid style={densely dotted,white!60!black}, 
%   xlabel=  $\omega  \,{[rad/s]}$, 		% x label
%   ylabel=  $\abs{W} {[\text{dB}]}$, 	% y label
%   legend style={at={(0.65,0.97)},anchor=north west},
%   legend cell align = {left},
%   xmin = 0.01, xmax = 100, ymin = -30, ymax = 30, % Hier kannst du die Achsenabschnitte definieren 
%        ]
%% DI        
%\addplot[blue, line width = 2, mark = o,  mark size =4pt] table[x expr = \thisrowno{0} ,y expr = \thisrowno{4} ,col sep=comma] {figures/E2P.csv};
%\addlegendentry{b = 1e-6} % Legendeneintrag

% We frequency shape

% Wu frequency shape

\begin{groupplot}[group style={
                      	group name=myplot,
                      	group size= 2 by 2,
                        vertical sep=2.25cm,
                        horizontal sep = 1.75cm},
                      	height=0.5\columnwidth,%0.33 for double column
                      	width = 0.48\columnwidth,%0.49 for double column
                      	xmajorgrids=true,
			ymajorgrids=true,
			 grid style={densely dotted,white!60!black},
			  xmin = -1, xmax = 25,
			  ymin = 0, ymax = 250,
			   ]
%%%%%%%%%%%%%%%%%%%%%%%% 11 %%%%%%%%%%%%%%%%%%%%%%%%%%%%%%
	\nextgroupplot[	title={$S$},
				 xmode=log,
				 xlabel= $\omega \,{[\text{rad/s}]}$,
				 ylabel=  $\abs{S} {[\text{dB}]}$, 
				xmin = 0.0032, xmax = 15,
			  	ymin = -20, ymax = 10,
				 ]
				 
\addplot[blue1, line width = 1.0,  no marks] table[x expr = \thisrowno{0} ,y expr = \thisrowno{1} ,col sep=comma] {figures/S.csv};\label{pl:0}
\addplot[blue2, line width = 1.0,  no marks] table[x expr = \thisrowno{0} ,y expr = \thisrowno{10} ,col sep=comma] {figures/S.csv};\label{pl:05pi}
\addplot[blue3, line width = 1.0,  no marks] table[x expr = \thisrowno{0} ,y expr = \thisrowno{19} ,col sep=comma] {figures/S.csv};\label{pl:pi}

\addplot[YellowOrange, line width = 1.25,  no marks] table[x expr = \thisrowno{0} ,y expr = \thisrowno{1} ,col sep=comma] {figures/We.csv};\label{pl:W0}
\addplot[purple, line width = 1.25,  no marks] table[x expr = \thisrowno{0} ,y expr = \thisrowno{2} ,col sep=comma] {figures/We.csv}; \label{pl:Wpi}

%%%%%%%%%%%%%%%%%%%%%%%% 21 %%%%%%%%%%%%%%%%%%%%%%%%%%%%%%
	\nextgroupplot[	title={$SP$},
				 xmode=log,
				 xlabel= $\omega \,{[\text{rad/s}]}$,
				ylabel=  $\abs{SP} {[\text{dB}]}$, 
				xmin = 0.0032, xmax = 15,
			  	ymin = 60, ymax = 130,
				 ]
          
\addplot[blue1, line width = 1.5,  no marks] table[x expr = \thisrowno{0} ,y expr = \thisrowno{1} ,col sep=comma] {figures/SP.csv};
\addplot[blue2, line width = 1.5,  no marks] table[x expr = \thisrowno{0} ,y expr = \thisrowno{10} ,col sep=comma] {figures/SP.csv};
\addplot[blue3, line width = 1.5,  no marks] table[x expr = \thisrowno{0} ,y expr = \thisrowno{19} ,col sep=comma] {figures/SP.csv};

\addplot[YellowOrange, line width = 1.25,  no marks] table[x expr = \thisrowno{0} ,y expr = \thisrowno{1} ,col sep=comma] {figures/VeVd.csv};
\addplot[purple, line width = 1.25,  no marks] table[x expr = \thisrowno{0} ,y expr = \thisrowno{2} ,col sep=comma] {figures/VeVd.csv};

%%%%%%%%%%%%%%%%%%%%%%%% 44 %%%%%%%%%%%%%%%%%%%%%%%%%%%%%%
	\nextgroupplot[	title={$KS$},
				 xmode=log,
				 xlabel= $\omega \,{[\text{rad/s}]}$,
				 ylabel=  $\abs{KS} {[\text{dB}]}$, 
				xmin = 0.0032, xmax = 15,
			  	ymin = -160, ymax = -70,
				 ]
          
\addplot[blue1, line width = 1.5,  no marks] table[x expr = \thisrowno{0} ,y expr = \thisrowno{1} ,col sep=comma] {figures/KS.csv};
\addplot[blue2, line width = 1.5,  no marks] table[x expr = \thisrowno{0} ,y expr = \thisrowno{10} ,col sep=comma] {figures/KS.csv};
\addplot[blue3, line width = 1.5,  no marks] table[x expr = \thisrowno{0} ,y expr = \thisrowno{19} ,col sep=comma] {figures/KS.csv};

\addplot[YellowOrange, line width = 1.25,  no marks] table[x expr = \thisrowno{0} ,y expr = \thisrowno{1} ,col sep=comma] {figures/VuVe.csv};
\addplot[purple, line width = 1.25,  no marks] table[x expr = \thisrowno{0} ,y expr = \thisrowno{2} ,col sep=comma] {figures/VuVe.csv};

%%%%%%%%%%%%%%%%%%%%%%%% 44 %%%%%%%%%%%%%%%%%%%%%%%%%%%%%%
	\nextgroupplot[	title={$KSP$},
				 xmode=log,
				 xlabel= $\omega \,{[\text{rad/s}]}$,
				 ylabel=  $\abs{KSP} {[\text{dB}]}$, 
				xmin = 0.0032, xmax = 15,
			  	ymin = -150, ymax = 40,
				 ]
          
\addplot[blue1, line width = 1.5,  no marks] table[x expr = \thisrowno{0} ,y expr = \thisrowno{1} ,col sep=comma] {figures/KSP.csv};
\addplot[blue2, line width = 1.5,  no marks] table[x expr = \thisrowno{0} ,y expr = \thisrowno{10} ,col sep=comma] {figures/KSP.csv};
\addplot[blue3, line width = 1.5,  no marks] table[x expr = \thisrowno{0} ,y expr = \thisrowno{19} ,col sep=comma] {figures/KSP.csv};

\addplot[YellowOrange, line width = 1.25,  no marks] table[x expr = \thisrowno{0} ,y expr = \thisrowno{1} ,col sep=comma] {figures/VuVd.csv};
\addplot[purple, line width = 1.25,  no marks] table[x expr = \thisrowno{0} ,y expr = \thisrowno{2} ,col sep=comma] {figures/VuVd.csv};

\end{groupplot}

\end{tikzpicture}
	\caption{Closed loop transfer functions ($0\,$deg: (\ref{pl:0}), $90\,$deg: (\ref{pl:05pi}), $180\,$deg: (\ref{pl:pi})) at selected pointing errors vs desired loop shape ($0\,$deg: (\ref{pl:W0}), $180\,$ deg: (\ref{pl:Wpi}).}
	\label{fig:BodeCtrl}
\end{figure}
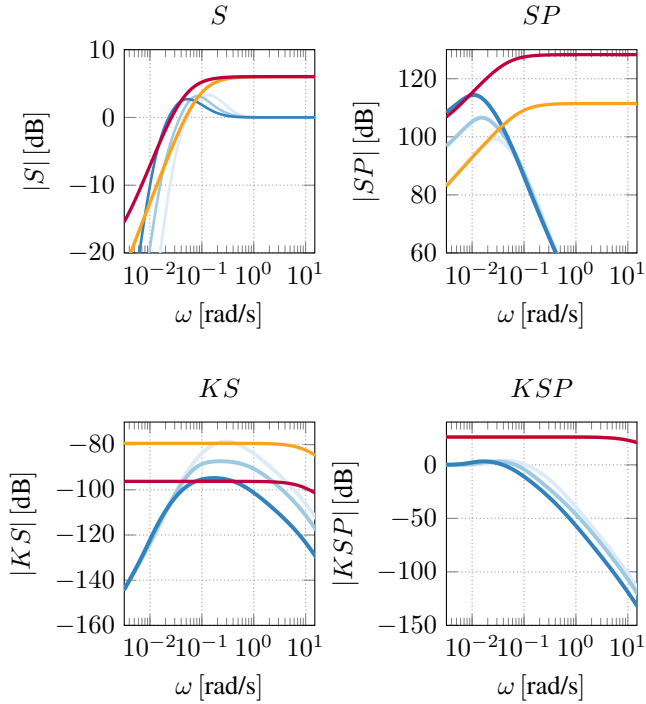

Fig.~\ref{fig:BodeCtrl} shows the resulting closed-loop transfer functions using the synthesized LPV controller at selected grid points ($0\,$deg: (\ref{pl:0}), $90\,$deg: (\ref{pl:05pi}), $180\,$deg: (\ref{pl:pi})). The figure also contains the weighting scheme (according to \eqref{eqn:mixedSens}) for pointing (\ref{pl:W0}) and start of the slew (\ref{pl:Wpi}). The comparison with the respective closed-loop transfer functions verifies that the calculated controller closely follows the imposed requirements.

\section{Controller Evaluation}\label{sec:Eval}
First, the LPV controllers stability margins are evaluated across the domain. Fig.~\ref{fig:Margins} shows the results. The smallest phase margin is $44.03\,$ deg. The smallest absolute gain margin is $13.15\,$ dB. Both margins comply well with the common aerospace requirement of $45\,$ deg, and $6\,$dB for phase and  gain margin, respectively.  
Generally, the gain and phase margins show an even distribution across the parameter domain. 
\begin{figure}[h!]
\centering
\begin{tikzpicture}
\definecolor{blue1}{RGB}{222,235,247}
\definecolor{blue2}{RGB}{158,202,225}
\definecolor{blue3}{RGB}{49,130,189}
%\begin{axis}[ width = 1.0\columnwidth, height = 0.5\columnwidth, % Grösse
%  	grid=major, 
%   grid style={densely dotted,white!60!black}, 
%   xlabel= Time after lift-off ${[\text{s}]}$, 		% x label
%   ylabel= Pitch Error ${[\text{deg}]}$, 	% y label
%   legend style={at={(0.5,0.01)},anchor=south east},
%   legend cell align = {left},
%   xmin = 25, xmax = 80, ymin = -0.9, ymax = 0.9, % Hier kannst du die Achsenabschnitte definieren 
%        

\begin{groupplot}[group style={
                      	group name=myplot,
                      	group size= 2 by 1,
                        vertical sep=2.25cm,
                        horizontal sep = 1.75cm},
                      	height=0.39\columnwidth,%0.33 for double column
                      	width = 0.48\columnwidth,%0.49 for double column
                      	xmajorgrids=true,
			ymajorgrids=true,
			 grid style={densely dotted,white!60!black},
			  xmin = 0, xmax = 180,
			  ymin = -30, ymax = 30,
			   ]]

\nextgroupplot[	title= Gain Margin,
				  ylabel= $\text{Gain Margin}  \,{[\text{dB}]}$,
				 xlabel= $\text{Pointing Error}  \,{[\text{deg}]}$,
				 ]
%%%%%%%%%% Gain Margin
\addplot[RoyalBlue, only marks, mark=x] table[x expr = \thisrowno{0} ,y expr = \thisrowno{2} ,col sep=comma] {figures/Margins.csv};\label{pl:Nom1}
\addplot[RoyalBlue, only marks, mark=x] table[x expr = \thisrowno{0} ,y expr = \thisrowno{3} ,col sep=comma] {figures/Margins.csv};\label{pl:Rob1}
\addplot[-, red!60, solid, line width = 1.5]coordinates{(0,6)(180,6)};\label{pl:req}
\addplot[-, red!60, solid, line width = 1.5]coordinates{(0,-6)(180,-6)};\label{pl:req}

\nextgroupplot[	title= Phase Margin,
				 ylabel= $\text{Phase Margin}  \,{[\text{deg}]}$,
				 xlabel= $\text{Pointing Error}  \,{[\text{deg}]}$,
			  	ymin = 0, ymax = 60,
				 ]
%%%%%%%%%% Phase margin
\addplot[RoyalBlue, only marks, mark=x] table[x expr = \thisrowno{0} ,y expr = \thisrowno{1} ,col sep=comma] {figures/Margins.csv};\label{pl:PM}
\addplot[-, red!60, solid, line width = 1.5]coordinates{(0,30)(180,30)};\label{pl:req}

\end{groupplot}
\end{tikzpicture}
\caption{Robustness margins of explicit controller along domain (\ref{pl:PM}) vs requirement (\ref{pl:req})}
\label{fig:Margins}
\end{figure}
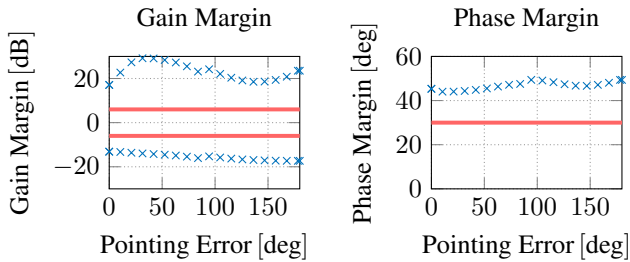
\begin{figure}[h!]
\centering
\begin{tikzpicture}
\definecolor{blue1}{RGB}{222,235,247}
\definecolor{blue2}{RGB}{158,202,225}
\definecolor{blue3}{RGB}{49,130,189}
%\begin{axis}[ width = 1.0\columnwidth, height = 0.5\columnwidth, % Grösse
%  	grid=major, 
%   grid style={densely dotted,white!60!black}, 
%   xlabel= Time after lift-off ${[\text{s}]}$, 		% x label
%   ylabel= Pitch Error ${[\text{deg}]}$, 	% y label
%   legend style={at={(0.5,0.01)},anchor=south east},
%   legend cell align = {left},
%   xmin = 25, xmax = 80, ymin = -0.9, ymax = 0.9, % Hier kannst du die Achsenabschnitte definieren 
%        

\begin{groupplot}[group style={
                      	group name=myplot,
                      	group size= 1 by 2,
                        vertical sep=2.25cm,
                        horizontal sep = 1.75cm},
                      	height=0.5\columnwidth,%0.33 for double column
                      	width = 0.9\columnwidth,%0.49 for double column
                      	xmajorgrids=true,
			ymajorgrids=true,
			 grid style={densely dotted,white!60!black},
			  xmin = 0, xmax = 500,
			  ymin = -35, ymax = 270,
			   ]]

\nextgroupplot[	title= Nominal Disturbance,
				  ylabel= $\theta  \,{[\text{deg}]}$,
				 xlabel= $\text{Time}  \,{[\text{deg}]}$,
				 ]
%%%%%%%%%% Nominal disturbance
\addplot[black, line width = 1.0] table[x expr = \thisrowno{0} ,y expr = \thisrowno{1} ,col sep=comma] {figures/RampResultsDist1dot0.csv};\label{pl:Nom1}
\addplot[Dandelion, line width = 1.0] table[x expr = \thisrowno{0} ,y expr = \thisrowno{2} ,col sep=comma] {figures/RampResultsDist1dot0.csv};\label{pl:Nom2}
\addplot[RoyalBlue, line width = 1.0] table[x expr = \thisrowno{0} ,y expr = \thisrowno{3} ,col sep=comma] {figures/RampResultsDist1dot0.csv};\label{pl:Nom3}
\addplot[Fuchsia, line width = 1.0] table[x expr = \thisrowno{0} ,y expr = \thisrowno{4} ,col sep=comma] {figures/RampResultsDist1dot0.csv};\label{pl:Nom4}

%\addplot[-, red!60, solid, line width = 1.5]coordinates{(0,-6)(180,-6)};\label{pl:req}

\nextgroupplot[	title= High Disturbance,
				 ylabel= $\theta  \,{[\text{deg}]}$,
				 xlabel= $\text{Time}  \,{[\text{s}]}$,
				 ]
%%%%%%%%%% Scaled disturbance
\addplot[black, line width = 1.0] table[x expr = \thisrowno{0} ,y expr = \thisrowno{1} ,col sep=comma] {figures/RampResultsDist1dot6.csv};\label{pl:Rob1}
\addplot[Dandelion, line width = 1.0] table[x expr = \thisrowno{0} ,y expr = \thisrowno{2} ,col sep=comma] {figures/RampResultsDist1dot6.csv};\label{pl:Rob2}
\addplot[RoyalBlue, line width = 1.0] table[x expr = \thisrowno{0} ,y expr = \thisrowno{3} ,col sep=comma] {figures/RampResultsDist1dot6.csv};\label{pl:Rob3}
\addplot[Fuchsia, line width = 1.0] table[x expr = \thisrowno{0} ,y expr = \thisrowno{4} ,col sep=comma] {figures/RampResultsDist1dot6.csv};\label{pl:Rob4}

\end{groupplot}
\end{tikzpicture}
\caption{Tracking of ramp reference signal (\ref{pl:Nom1}): Pointing $H_\infty$ (\ref{pl:Nom2}),  LPV (\ref{pl:Nom3}), Discrete Switching (\ref{pl:Nom4})}
\label{fig:Results}
\end{figure}
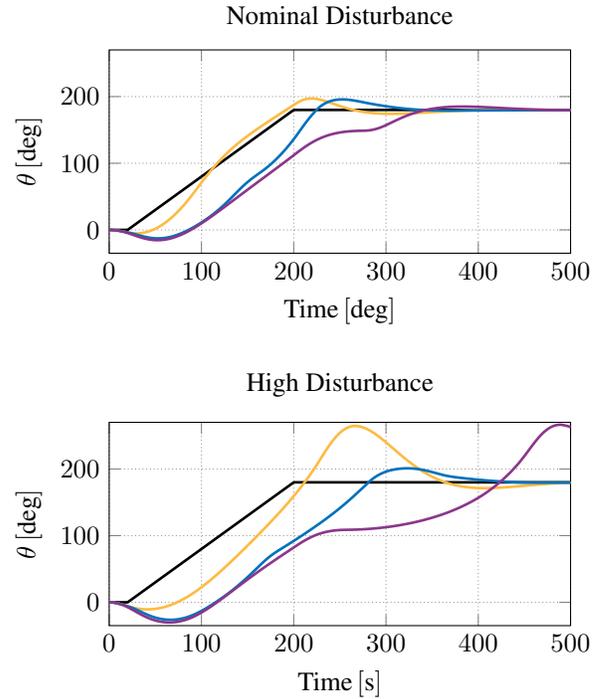
Time domain simulations are conducted in Matlab/Simulink to investigate the qualitative behaviour of the LPV controller. 
Here, the CubeSat shall perform a $180\,$deg slew by tracking a ramp signal with a slope of $1\,$ deg/s starting at $20\,$s. Afterwards, the final attitude of $180\,$deg shall be maintained. The closed loop system is subject to the MagLev specific disturbance torque described in section~\ref{sec:CtrlProb}. Fig.~\ref{fig:Results} shows the simulation results.
The closed loop follows the ramp after a short delay with a relatively constant offset, which reflects the design goals. As the CubeSat approaches the desired attitude, it starts to converge to the ramp signal. This behavior
is expected as the controller gains get biased towards the pointing requirements. The CubeSat first intercepts the desired $180\,$ degrees after $225\,$s, converges inside $1\,$deg after
$225\,$s, and remains inside $0.1\,$ deg after $495\,$s. The maximum overshot is approximately $8\%$.
Commonly, CubeSats are only equipped with a single controller. Most often, a classical fix gain PID controller tuned for a mission specific pointing task or limited to a compromise between pointing and slewing. Thus, the controller cannot
autonomously adapt to differently sized and shaped slew maneuvers, i.e., smaller or larger changes in attitude. To show the advantages of the
LPV controller, two classical $H_\infty$ controller are designed. The first controller uses the pointing weights, while the second uses the slewing weights.  
The simulation results of the pointing controller (\ref{pl:Nom2}) are shown in Fig.~\ref{fig:Results}. The controller tracks the ramp closer given its design goals. However, this is
also due to the fact that the controller is in saturation for the first $65\,$s of the slew. Note that in this case the controller cannot counteract any further disturbances.
The controller reaches the desired attitude after $191\,$s and overshoots it by approximately $10\%$.  The pointing controller converges inside $1\,$deg after
$384\,$s and remains inside $0.2\,$ deg after $520\,$s. Although, initially acquiring the attitude faster the pointing controller requires longer to converge. Although not pictured: Note that the slewing controller failed to reach the final attitude.
An, alternative to a single controller is presented by discretely switching between a dedicated slewing and pointing controller as it is common for larger satellites. Here, a simple switching scheme (\cite{Geshnizjani2013Thesis}) for agile satellites is implemented using the two synthesized $H_\infty$ controllers. The simulation results are shown in Fig.~\ref{fig:Results}. This approach performs worse than the other two controllers.

All three presented control approaches are now evaluated for a disturbance which is scaled by a factor $1.6$ to evaluate the robustness of the approaches.
The bottom plot in Fig.~\ref{fig:Results} presents the results. It can be seen that the discrete switching fails to acquire the desired attitude. The pointing $H_\infty$ controller shows significantly degraded performance. Most, notably the overshot increases to approximately $36\%$ as the thrusters are in saturation from $22\,$s until $224\,$s. The $1\,$ deg band is reached after $490\,$s. The LPV controller demonstrates very good robustness. The overshoot only increases to $11\%$ as the controller is sporadically in saturation (total of $60\,$s) during the slew and shortly during the transition. The $1\,$deg error band is reached after $426\,$s. 

This investigation demonstrates the advantages of a single LPV controller covering both slewing and pointing maneuvers. The approach provides larger robustness than the single fixed
gain controllers or discretely switched controllers.

%\begin{figure}[h!]elatively
%\centering
%%\input{figures/Results}
%\input{figures/BangResults}
%\caption{Tracking of "bang bang" reference signal (\ref{pl:Nom1}): Pointing $H_\infty$ (\ref{pl:Nom2}),  LPV Controller (\ref{pl:Nom3}), Discrete Switching (\ref{pl:Nom4})}
%\label{fig:Results}
%\end{figure}
%
%\begin{figure}[h!]
%\centering
%%\input{figures/Results}
%\input{figures/StepResults}
%\caption{Tracking of step reference signal (\ref{pl:Nom1}): Pointing $H_\infty$ (\ref{pl:Nom2}),  LPV Controller (\ref{pl:Nom3}), Discrete Switching (\ref{pl:Nom4})}
%\label{fig:Results}
%\end{figure}

\section{Conclusions}
The present paper demonstrates the advantages of designing a single linear parameter-varying (LPV) controller for the attitude control of CubeSats using electrospray thrusters. 
The approach allows to explicitly account for different control requirements of the slewing and pointing segments of larger attitude maneuvers. Thus, apparent thrust limitations can be considered
to avoid controller saturations, which allow for more robust controllers.
The performance and robustness of the controller is demonstrated on a simulation of MIT Space Propulsion Lab's Magnetic Levitation CubeSat Testbed.
Future work includes the extension of the controller architecture with an anti-windup scheme using a novel structure LPV synthesis, as well as 
hardware tests using the MagLev device.

\acknowledgments
This work was partially supported by the European Union under Grant No. 101153910 entitled ''Recommissioning and Deorbiting using Cubesat Swarms with Electro Spray Thrusters''. Views and opinions expressed are however those of the authors only and do not necessarily reflect those of the European Union. Neither the European Union nor the granting authority can be held responsible for them.

%%%%%%%%%%%%%%%%%%%%%%%%%%%%%%%%%%%%%%%%%%%%%%%%%%%%%%%%%%%%%%%%%%%%%%%%%%%%%%%%%%%%%%%%%%%%%%%%%%%%%%
\bibliographystyle{IEEEtran}
\bibliography{iEPS}

%\begin{thebibliography}{1}
%
%\bibitem{ITAR}
%U.S. Munitions List, Sections 38 and 47(7) of the Arms Export Control Act (22 U.S.C 2778 and 2794(7).
%
%\bibitem{AeroConf}
%
%\end{thebibliography}

%%%%%%%%%%%%%%%%%%%%%%%%%%%%%%%%%%%%%%%%%%%%%%%%%%%%%%%%%%%%%%%%%%%%%%%%%%%%%%%%%%%%%%%%%%%%%%%%%%%%%%
\thebiography
%% This biostyle allows you to insert your photo size 1in X 1.25in
\begin{biographywithpic}
{Felix Biertümpfel}{figures/FBiertuempfel.jpg}
received his PhD degree from the University of Nottingham, United Kingdom, in 2021. Since 2021 he works as a research associate at the Technische Universit\"at Dresden at the Chair of Flight Mechanics and Control. As a Marie-Curie Fellow, he works on novel approaches for space debris removal together with the MIT and the University of Michigan. His research interest is in robust and linear time-varying control for space applications.
\end{biographywithpic}

\begin{biographywithpic}{Emily Burgin}{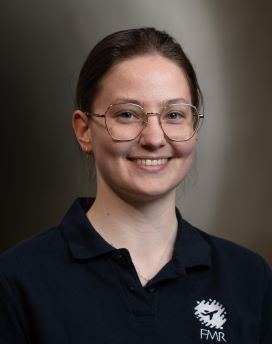}
	is currently a doctoral candidate at the Technische Universit\"at Dresden, Germany, at the Chair of Flight Mechanics and Control. The focus of her research is in linear parameter-varying techniques to aerospace systems. She received an MEng in Aerospace Engineering from the University of Nottingham, England, in 2021. Since then, she worked as a GNC engineer in at Deimos Space UK before pursuing a PhD.
\end{biographywithpic}

\begin{biographywithpic}
{Harald Pfifer}{figures/HPfifer00.jpg}
holds the Chair in Flight
	Mechanics and Control at Technische Universit\"at Dresden, Germany. He received his Ph.D.
	from Technical University Munich, Germany, in 2013. Before joining Technische Universit\"at
	Dresden in 2021, he was an assistant professor at University of Nottingham, and a post-doctoral
	associate at University of Minnesota. His main research interests include aeroservoelastic control, uncertain dynamical systems, and robust and linear parameter-varying control.
\end{biographywithpic}

\begin{biographywithpic}
{Hanna-Lee Harjono}{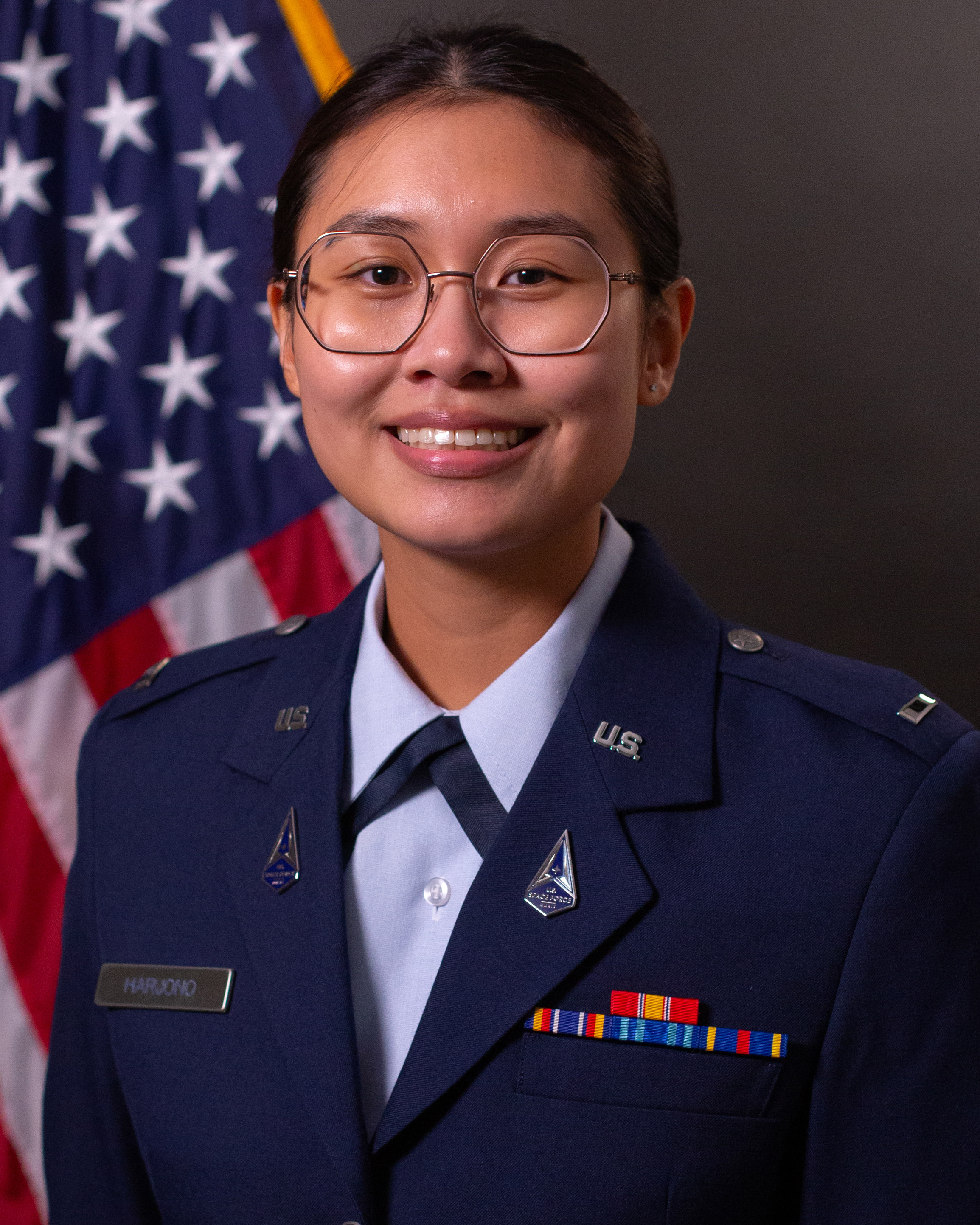}
is currently a 1st Lieutenant in the US Space Force as a Developmental Engineer. She received an M.S. in Aeronautics and Astronautics from the Massachusetts Institute of Technology in 2023. The focus of her research was on the development of control algorithms for electrospray propulsion systems. Post-graduation, she is working as a Mission Operations Flight Commander with the National Reconnaisance Office.
\end{biographywithpic}

\begin{biographywithpic}
{Paulo Lozano}{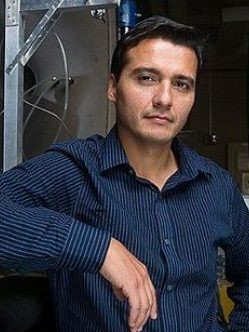}
holds a PhD in Aerospace Engineering from the Massachusetts Institute of Technology and is currently the M. Aleman Velasco Professor of Aerospace Engineering. He is director of the Space Propulsion Lab at MIT and Head of the Space Sector. His research interests include space propulsion, electrospray thrusters, micro- and nano-fabrication, space mission design, small satellite technology development, ion beams and micro-fluidics.\end{biographywithpic}

\end{document}